\documentclass[twocolumn,showpacs,preprintnumbers,amsmath,amssymb]{revtex4}
\usepackage{dcolumn}
\usepackage{bm}
\usepackage{graphicx}

\def\wco{{\omega_{c}}}
\def\w0{\omega_0}
\def\k0{k_0}

\newcommand{\bkj}{b_{\mathbf{k}_dj}}
\def\bbm[#1]{\mbox{\boldmath $#1$}}
\newcommand{\Ert}{\mathbf{E}(\mathbf{r},t)}

\newcommand{\Brt}{\mathbf{B}(\mathbf{r},t)}

\newcommand{\akj}{a_{\mathbf{k}j}}
\newcommand{\ackj}{a_{\mathbf{k}j}^\dag}

\newcommand{\ackjp}{a_{\mathbf{k}'j'}^\dag}

\newcommand{\kx}{k_x}
\newcommand{\ky}{k_y}
\newcommand{\kz}{k_z}
\newcommand{\kp}{k_\parallel}
\newcommand{\kdz}{k_{dz}}
\newcommand{\fRjr}{\mathbf{f}_R\bigl(\mathbf{k}j,\mathbf{r}\bigr)}
\newcommand{\fsRjr}{\mathbf{f}^*_R\bigl(\mathbf{k}j,\mathbf{r}\bigr)}
\newcommand{\fRunor}{\mathbf{f}_R\bigl(\mathbf{k}1,\mathbf{r}\bigr)}
\newcommand{\fRduer}{\mathbf{f}_R\bigl(\mathbf{k}2,\mathbf{r}\bigr)}
\newcommand{\fLjr}{\mathbf{f}_L\bigl(\mathbf{k}_dj,\mathbf{r}\bigr)}
\newcommand{\fsLjr}{\mathbf{f}^*_L\bigl(\mathbf{k}_dj,\mathbf{r}\bigr)}
\newcommand{\fLunor}{\mathbf{f}_L\bigl(\mathbf{k}_d1,\mathbf{r}\bigr)}
\newcommand{\fLduer}{\mathbf{f}_L\bigl(\mathbf{k}_d2,\mathbf{r}\bigr)}

\newcommand{\fRunorm}{\mathbf{f}_R^>\bigl(\mathbf{k}1,\mathbf{r}\bigr)}
\newcommand{\fRduerm}{\mathbf{f}_R^>\bigl(\mathbf{k}2,\mathbf{r}\bigr)}

\newcommand{\fLunorm}{\mathbf{f}_L^>\bigl(\mathbf{k}_d1,\mathbf{r}\bigr)}
\newcommand{\fLduerm}{\mathbf{f}_L^>\bigl(\mathbf{k}_d2,\mathbf{r}\bigr)}
\newcommand{\bckj}{b_{\mathbf{k}_dj}^\dag}
\newcommand{\bckjp}{b_{\mathbf{k}_d'j'}^\dag}
\newcommand{\Equadroeta}{\langle\mathbf{E}^2\rangle_{\eta\ \mathrm{R}}^{\mathrm{con}}}

\newcommand{\Equadrolim}{\langle\mathbf{E}^2\rangle_{\mathrm{R}}^{\mathrm{con}}}

\begin{document}

\title{Electromagnetic field fluctuations near a dielectric-vacuum boundary and surface divergences in the ideal conductor limit}

\author{Nicola Bartolo\mbox{${\ }^{1}$} and Roberto Passante\mbox{${\ }^{2}$}}
\affiliation{\mbox{${\ }^{1}$} INO-CNR BEC Center and Dipartimento di Fisica, Universit\`{a} di Trento, Via Sommarive 14, I-38123 Povo, Trento, Italy
\\
\mbox{${\ }^{2}$}Dipartimento di Fisica dell'Universit\`{a} degli Studi di Palermo and CNISM, Via Archirafi 36, I-90123 Palermo, Italy}

\email{roberto.passante@unipa.it}


\begin{abstract}
We consider the electric and magnetic field fluctuations in the vacuum state in the region external to a half-space filled with a homogeneous non-dissipative dielectric. We discuss  an appropriate limit to an ideal metal and concentrate our interest on the renormalized field fluctuations, or equivalently to renormalized electric and magnetic energy densities, in the proximity of the dielectric-vacuum interface. We show that surface divergences of field fluctuations arise at the interface in an appropriate ideal conductor limit, and that our limiting procedure allows to discuss in detail their structure. Field fluctuations close to the surface can be investigated through the retarded Casimir-Polder interaction with an appropriate polarizable body.
\end{abstract}

\maketitle

\section{\label{sec:1}Introduction}

Vacuum fluctuations are an outstanding consequence of the quantum theory of the electromagnetic radiation field, and observable manifestations of vacuum fluctuations include radiative level shifts and Casimir forces \cite{Milonni94,CPP95}. These forces are long-range interactions between neutral objects due to their interaction with vacuum fluctuations; they are quantum effects of the radiation field and have not a classical analogue. Usually Casimir effect refers with this kind of interactions between macroscopic objects \cite{Casimir48}, and Casimir-Polder forces with the interactions of atoms or molecules with a macroscopic object such as a surface or among atoms and molecules \cite{CP48}. Vacuum fluctuations are changed by the presence of boundary conditions given by dielectric or metallic surfaces. This change of electric and magnetic field fluctuations gives rise to energy shifts of atoms or molecules, placed near the surface, that depend from their position, yielding Casimir-Polder forces \cite{MPRSV08}. Evaluating field fluctuations around a boundary allows also to obtain how interactomic Casimir-Polder forces are modified by the presence of the boundary \cite{SPR06,PS07}. These facts give further elements making relevant studying the dependence of vacuum fluctuations or vacuum energy densities on the distance from a dielectric or conducting surface.

In the case of a perfectly conducting infinite plate, the vacuum electric and magnetic fluctuations $\langle E^2 \rangle$ and $\langle B^2 \rangle$, after subtraction of the homogeneous zero-point terms existing even in the absence of the plate, behave as $z^{-4}$, $z$ being the distance from the plate; they do diverge at the metal-vacuum interface. The physical origin of this divergence and the possible presence of further singular terms at the interface has been recently questioned in the literature \cite{MCPW06,Milton11}, and possibility of their regularizations has been investigated in the case of a scalar field using an appropriate potential to represent the wall \cite{Milton11a}. It has been shown that, in the case of a half-space filled with a non-dissipative dielectric material characterized by a real dielectric constant $\epsilon$ independent from the frequency, such divergences at the interface are still present in the limit $\epsilon \to \infty$ \cite{HL99}. An analogous behavior has been obtained also if dispersion is introduced using the plasma model \cite{SF02}. In the case of a scalar field, it has been shown that the stress-tensor components are regularized by reflection and transmission coefficients; also, the surface divergence associated to a perfectly reflecting mirror is canceled by a new energy density near the surface \cite{Pfenning00}. Surface divergences in the Casimir energies has also been considered by introducing extra terms in the Lagrangian in the form of a $\delta$-function potential, in order to simulate the boundary condition \cite{Milton04}. Research in this direction has been also motivated by the quest for situations where negative energy densities occur \cite{SF05}, and solving discrepancies between total self-energies and the local energy densities \cite{MCPW06,FS98}. The presence of surface divergences may be relevant also from the point of view of the coupling of the related field energy densities with gravity \cite{Milton11,MNS11}.

In this paper we shall consider electric and magnetic field fluctuations in the vacuum region near a half-space filled with a homogeneous nondissipative dielectric characterized by a real refractive index $n$. The field fluctuations are evaluated using the Carniglia-Mandel triplets as field modes. A time-splitting procedure is used in evaluating the frequency integrals, and we notice that this is mathematically equivalent to introduce an exponential cut-off function  $e^{-\eta \omega}$, with $1/\eta=\wco$ playing the role of a cut-off frequency. We then consider the limit $n \to \infty$, leading to a model for a metallic material. If the quantity $1/\eta$ introduced by the time-splitting procedure is kept finite, this is representative of a situation where the contribution of field modes with $\omega > \wco$ is suppressed by the exponential cut-off. We can then consider the limit of an ideal metallic plate by taking the limit $1/\eta=\wco \to \infty$. After this limit, the cut-off function is one for all frequencies and we recover the nondispersive case. With this procedure, the ideal metal case is obtained through a limiting procedure, and at no point we need to use boundary conditions for ideal metals. We find that in the limit $n \to \infty$ the renormalized field fluctuations are finite for any finite value of $\wco$. We are also able to show that in the ideal metal limit $\wco \to \infty$ surface divergences at the metal-vacuum interface emerge, and to investigate their properties. As far as we know, the structure of these surface divergences cannot be derived if the field is directly quantized with a perfectly conducting boundary. Moreover, no discrepancy between total self-energy and local energy densities is present in our model. Such discrepancies indeed occur when energy densities are evaluated directly for the ideal metallic plate, because of the singular behavior at $z=0$. From the field energy density we can also evaluate the electric and magnetic Casimir-Polder force on an atom placed near the dielectric half-space. Strong radiative interactions of cesium atoms near dielectric boundaries have been recently investigated experimentally \cite{ASALOVK11}.

In Sec. \ref{sec:2} we calculate the vacuum fluctuations of the electric and magnetic field in the vacuum space in presence of a nondissipative homogeneous dielectric half-space. We use quantization in terms of the Carniglia-Mandel triplets, that include evanescent waves, and we use the known time-splitting procedure. In Sec. \ref{sec:3} we consider the ideal metal case by a limit procedure from the dielectric case, and we analyze in detail the field fluctuations (and field energy densities), in particular in proximity of the metal-vacuum interface. The emergence and structure of surface divergences at the interface in the ideal conductor limit is discussed in detail.

\section{\label{sec:2}Quantum fluctuations of the electromagnetic field near the dielectric-vacuum interface}

We consider an half-space filled with an homogeneous dielectric medium and the vacuum in the other half-space. We label $z$ the direction orthogonal to the dielectric-vacuum interface: $z<0$ is the dielectric half-space and $z>0$ the vacuum half-space. Our model has a translational symmetry in the x-y directions. We assume that the dielectric is non-dissipative with a (real) dielectric constant independent on the frequency. We shall later discuss how to introduce in our model a dependence on the frequency, in particular when considering the limit for an ideal conductor.

First step is the calculation of the square of the electric and magnetic fields in the vacuum half-space ($z>0$) in the ground state of the field. We use the well-known quantization scheme in terms of the Carniglia-Mandel triplets \cite{CM71}. The Carniglia-Mandel modes are given by the following expressions

\begin{widetext}
\begin{subequations}\label{TriplettiCM}
\begin{equation}\label{fR1}
\shoveleft\qquad\qquad
\fRunor=\frac{\hat{\mathbf{e}}_1}{(2\pi)^{3/2}}
\begin{cases}
 \frac{2\kz}{\kz+\kdz}\ e^{i\mathbf{k}_d^-\cdot\mathbf{r}} &z<0\\
 e^{i\mathbf{k}^-\cdot\mathbf{r}} + \frac{\kz-\kdz}{\kz+\kdz}\ e^{i\mathbf{k}^+\cdot\mathbf{r}} &z\ge0\\
\end{cases}
\end{equation}
\begin{equation}\label{fR2}
\shoveleft\qquad\qquad
\fRduer=\frac{\hat{\mathbf{e}}_2}{(2\pi)^{3/2}}
\begin{cases}
 \frac{2n\kz}{n^2\kz+\kdz}\ e^{i\mathbf{k}_d^-\cdot\mathbf{r}} &z<0\\
 e^{i\mathbf{k}^-\cdot\mathbf{r}} + \frac{n^2\kz-\kdz}{n^2\kz+\kdz}\ e^{i\mathbf{k}^+\cdot\mathbf{r}} &z\ge0\\
\end{cases}
\end{equation}
\begin{equation}\label{fL1}
\shoveleft\qquad\qquad
\fLunor=\frac{\hat{\mathbf{e}}_1}{(2\pi)^{3/2}}\frac{1}{n}
\begin{cases}
 e^{i\mathbf{k}_d^+\cdot\mathbf{r}} + \frac{\kdz-\kz}{\kdz+\kz}\ e^{i\mathbf{k}_d^-\cdot\mathbf{r}} &z<0\\
 \frac{2\kdz}{\kdz+\kz}\ e^{i\mathbf{k}^+\cdot\mathbf{r}} &z\ge0\\
\end{cases}
\end{equation}
\begin{equation}\label{fL2}
\shoveleft\qquad\qquad
\fLduer=\frac{\hat{\mathbf{e}}_2}{(2\pi)^{3/2}}\frac{1}{n}
\begin{cases}
 e^{i\mathbf{k}_d^+\cdot\mathbf{r}} + \frac{\kdz-n^2\kz}{\kdz+n^2\kz}\ e^{i\mathbf{k}_d^-\cdot\mathbf{r}} &z<0\\
 \frac{2n\kdz}{\kdz+n^2\kz}\ e^{i\mathbf{k}^+\cdot\mathbf{r}} &z\ge0.\\
\end{cases}
\end{equation}
\end{subequations}
\end{widetext}
where the index $L,R$ indicates waves propagating toward the dielectric-vacuum interface from left or right, respectively, and $j=1$ specifies transverse electric (TE) modes and $j=2$ transverse magnetic (TM) modes. We have also defined the polarization operators
\begin{equation}\label{PolarizzazioneDielettrico}
\begin{aligned}
 \hat{\mathbf{e}}_1=&(-\Delta_\parallel)^{-1/2} \bigl( -i\partial_y;i\partial_x;0 \bigr)\\
 \hat{\mathbf{e}}_2=&(\Delta\Delta_\parallel)^{-1/2} \bigl( -\partial_x\partial_z;-\partial_y\partial_z;\Delta_\parallel \bigr)\\
\end{aligned}
\end{equation}
where $\Delta=\partial_x^2+\partial_y^2+\partial_z^2$ and $\Delta_\parallel=\partial_x^2+\partial_y^2$. These operators, acting on the plane-wave parts of
\eqref{TriplettiCM}, give the appropriate polarization unit vectors.

In order to distinguish waves propagating in the positive and negative directions of the $z$ axis, we have defined in the free space ($z>0$) the wavevectors
\begin{equation}
 \mathbf{k}^\pm=(k_x;k_y;\pm k_z)=(\mathbf{k}_\parallel;\pm k_z).
\end{equation}
Inside the dielectric ($z<0$) the wavevectors are
\begin{equation}
 \mathbf{k}_d^\pm=(\mathbf{k}_\parallel;\pm\kdz),
\end{equation}
with
\begin{equation}
\begin{aligned}
\label{KappaZetaD}
 \kdz=&\sqrt{(n^2-1)\kp^2+n^2\kz^2} \\
 \kz=&\frac{1}{n}\sqrt{\kdz^2-(n^2-1)\kp^2} \\
 \end{aligned}
\end{equation}

At the dielectric-vacuum interface ($z=0$), $\mathbf{k}_\parallel$ and the frequency $\omega_k$ are conserved, while $k_z$  appropriately changes in order to ensure the required continuity of $\mathbf{E}_\parallel$, $\mathbf{D}_\perp$ and $\mathbf{B}$. For $\kdz^2$ less than $(n^2-1)\kp^2$, $\kz$ becomes imaginary and a total internal reflection of L-modes occurs at the interface yielding an evanescent wave in the vacuum region. Thus evanescent waves are correctly taken into account by using the Carniglia-Mandel modes. Proof of  orthogonality and completeness of these modes can be found in \cite{CM71,BBB72}. Similar modes have recently been used for the quantization of the electromagnetic field in the presence of a nondispersive and nondissipative dielectric slab \cite{CRE09,CRE09b}.

The explicit expression of the modes for $z>0$ (vacuum region) are

\begin{widetext}
\begin{subequations}\label{TriplettiZetaPositivo}
\begin{equation}
\shoveleft
 \fRunorm=\frac{1}{(2\pi)^{3/2}}\ \frac{1}{\kp}\bigl(\ky;-\kx;0\bigr)\ \Biggl[e^{i\mathbf{k}^-\cdot\mathbf{r}} + \frac{\kz-\kdz}{\kz+\kdz}e^{i\mathbf{k}^+\cdot\mathbf{r}}\Biggr];
\end{equation}
\begin{equation}
\begin{aligned}
 \fRduerm=\frac{1}{(2\pi)^{3/2}}\ \Biggl[\frac{1}{k\kp}\bigl(\kx\kz;\ky\kz;\kp^2\bigr)\ e^{i\mathbf{k}^-\cdot\mathbf{r}}\qquad\qquad\qquad\qquad\qquad\quad\\
           -\frac{1}{k\kp}\bigl(\kx\kz;\ky\kz;-\kp^2\bigr)\  \frac{n^2\kz-\kdz}{n^2\kz+\kdz}e^{i\mathbf{k}^+\cdot\mathbf{r}}\Biggr];
\end{aligned}
\end{equation}
\begin{equation}
\shoveleft\fLunorm=\frac{1}{(2\pi)^{3/2}}\ \frac{1}{n\kp}\bigl(\ky;-\kx;0\bigr)\ \frac{2\kdz}{\kdz+\kz} e^{i\mathbf{k}^+\cdot\mathbf{r}};\qquad\qquad\qquad
\end{equation}
\begin{equation}
\shoveleft\fLduerm=-\frac{1}{(2\pi)^{3/2}}\ \frac 1{nk\kp}\bigl(\kx\kz;\ky\kz;-\kp^2\bigr)\ \frac{2n\kdz}{\kdz+n^2\kz} e^{i\mathbf{k}^+\cdot\mathbf{r}}.
\end{equation}
\end{subequations}

In terms of the modes \eqref{TriplettiCM}, we can write the expressions of the electric and magnetic field operators

\begin{equation}\label{EspansioneETripletti}
\begin{split}
 \Ert=\sum_j\int_{\kz>0}\!\!\!\!\!\!\!\!d^3\mathbf{k}\ i\sqrt{2\pi\hbar\omega_k}\ \biggl[\akj\ e^{-i\omega_k t}\ \fRjr - \ackj e^{i\omega_k t}\ \fsRjr\biggr]\\
     +\sum_j\int_{\kdz>0}\!\!\!\!\!\!\!\!\!d^3\mathbf{k}_d\ i\sqrt{2\pi\hbar\omega_k}\ \biggl[\bkj\ e^{-i\omega_k t}\ \fLjr - \bckj e^{i\omega_k t}\ \fsLjr\biggr];
\end{split}
\end{equation}
\begin{equation}\label{EspansioneBTripletti}
\begin{split}
 \Brt=\sum_j\int_{\kz>0}\!\!\!\!\!\!\!\!d^3\mathbf{k}\ \sqrt{\frac{2\pi\hbar c^2}{\omega_k}}\ \biggl[\akj\ e^{-i\omega_k t}\ \nabla\times\fRjr + \ackj e^{i\omega_k t}\ \nabla\times\fsRjr\biggr]\\
     +\sum_j\int_{\kdz>0}\!\!\!\!\!\!\!\!\!d^3\mathbf{k}_d\ \sqrt{\frac{2\pi\hbar c^2}{\omega_k}}\ \biggl[\bkj\ e^{-i\omega_k t}\ \nabla\times\fLjr + \bckj e^{i\omega_k t}\ \nabla\times\fsLjr\biggr].
\end{split}
\end{equation}
(operators $\akj$, $\ackj$ and $\bkj$, $\bckj$ refer to R- and L-modes, respectively).

The field Hamiltonian, after subtraction of the zero-point energy, is given by
\begin{equation}\label{EnergiaDielettrico2}
 H_F=\sum_j\int_{\kz>0}\!\!\!\!\!\!d^3\mathbf{k}\ \hbar\omega_k\ \ackj\akj \\
    +\sum_j\int_{\kdz>0}\!\!\!\!\!\!\!d^3\mathbf{k}_d\ \hbar\omega_k\ \bckj\bkj ,
\end{equation}
\end{widetext}
where the annihilation and creation operators for photons in the different field modes satisfy the usual bosonic commutation rules
\begin{equation}
\begin{aligned}
 \bigl[\akj,\ackjp\bigr]=&\ \delta_{jj'}\ \delta^3(\mathbf{k}-\mathbf{k}')\\
 \bigl[\bkj,\bckjp\bigr]=&\ \delta_{jj'}\ \delta^3(\mathbf{k}_d-\mathbf{k}_d')\\
\end{aligned}
\end{equation}
(all other commutators vanish).

Using the expression \eqref{EspansioneETripletti} for the electric field with the modes \eqref{TriplettiZetaPositivo}, we can evaluate the average value of a space-time correlation of the electric field in the vacuum space near the interface $(z>0)$ on the ground state of the field, obtaining (this calculations is similar to that in \cite{HL99}, and we are giving some detail of it in order to show how the time-splitting procedure we are going to use is equivalent to a high-frequency cut-off)
\begin{equation}
\begin{aligned}
&\langle\mathrm{E}_\lambda(\mathbf{r},t)\mathrm{E}_\lambda(\mathbf{r}',t')\rangle \\
&=2\pi\hbar \Big[\sum_{j} \int_{\kz>0}\!\!\!\!\!d^3\mathbf{k}\ \omega_k\ e^{-i\omega_k(t-t')}\ \mathrm{f}_{R\lambda}^{>}\bigl(\mathbf{k}j,\mathbf{r}\bigr)\ \mathrm{f}_{R\lambda}^{>*}\bigl(\mathbf{k}j,\mathbf{r}'\bigr)\\
&+\sum_{j} \int_{\kdz>0}\!\!\!\!\!\!d^3\mathbf{k}_d\ \omega_k\ e^{-i\omega_k(t-t')}\ \mathrm{f}_{L\lambda}^{>}\bigl(\mathbf{k}_dj,\mathbf{r}\bigr)\ \mathrm{f}_{L\lambda}^{>*}\bigl(\mathbf{k}_dj,\mathbf{r}'\bigr)\Big]\\
\end{aligned}
\end{equation}
with $\lambda = x,y,z$. This quantity diverges for $\mathbf{r}'\to\mathbf{r}$ and $t'\to t$, but we can use a point-splitting procedure, by introducing the following quantity
\begin{equation}
\langle \mathrm{E}_\lambda^2\rangle_\eta=\langle \mathrm{E}_\lambda(\mathbf{r},t)\,\mathrm{E}_\lambda(\mathbf{r},t'=t+i\eta)\rangle
\end{equation}
with $\eta >0$.

In the next section we shall show that this time-splitting procedure is mathematically equivalent to introduce a cut-off frequency in the frequency integrals, and this allows us to obtain the ideal conductor case as a limit process.
Thus, for $\eta\to 0$ we have $\langle \mathrm{E}_\lambda^2\rangle_\eta\to\langle \mathrm{E}_\lambda^2\rangle$. As long as $\eta$ is finite, the integrals do not diverge and we obtain
\begin{equation}\label{ContributoParziale1}
\begin{aligned}
\langle\mathrm{E}_\lambda^2\rangle_\eta=
2\pi\hbar \sum_{j} &\int_{\kz>0}\!\!\!\!\!d^3\mathbf{k}\ \omega_k\ e^{-\eta\omega_k}\ |\mathrm{f}_{R\lambda}^{>}\bigl(\mathbf{k}j,\mathbf{r}\bigr)|^2\\
+2\pi\hbar \sum_{j} &\int_{\kdz>0}\!\!\!\!\!\!\!d^3\mathbf{k}_d\ \omega_k\ e^{-\eta\omega_k}\ |\mathrm{f}_{L\lambda}^{>}\bigl(\mathbf{k}_dj,\mathbf{r}\bigr)|^2.\\
\end{aligned}
\end{equation}
In this expression, contributions from both traveling and evanescent waves are taken into account.

After some algebraic calculations, using \eqref{KappaZetaD} and \eqref{TriplettiZetaPositivo}, we obtain the following integral expression for $\langle \mathbf{E}^2\rangle_\eta$ for $z>0$ and arbitrary $n$

\begin{widetext}
\begin{equation}\label{EQuadroMedio}
\begin{aligned}
\langle \mathbf{E}^2\rangle_\eta
&=\frac{\hbar c}{2\pi} \int_0^\infty\!\!\!\!\!d\kp\int_0^\infty\!\!\!\!\!d\kz\ \kp k \left\{ 2+\Bigl(\frac{\kz-\kdz}{\kz+\kdz}\Bigl)^2+\Bigl(\frac{n^2\kz-\kdz}{n^2\kz+\kdz}\Bigl)^2\right.\\
&+2\frac{\kz-\kdz}{\kz+\kdz}\cos\bigl(2\kz z\bigr)+2\Bigl(2\frac{\kp^2}{k^2}-1\Bigr)\frac{n^2\kz-\kdz}{n^2\kz+\kdz}\cos\bigl(2\kz z\bigr)
+\frac{\kz}{\kdz}\left[ \Bigl(\frac{2\kdz}{\kdz+\kz}\Bigr)^2
+\Bigl(\frac{2n\kdz}{\kdz+n^2\kz}\Bigr)^2\ \right]\\
&\left. +\frac{|\kz|}{\kdz} \left[\frac{4\kdz^2}{\kdz^2+|\kz|^2}
+ \frac{4n^2\kdz^2}{\kdz^2+n^4|\kz|^2}\right] e^{-2|\kz|z}\right\} e^{-\eta ck}.
\end{aligned}
\end{equation}

An analogous expression can be obtained for the magnetic part ($z>0$ and arbitrary $n$)

\begin{equation}\label{BQuadroMedio}
\begin{aligned}
\langle \mathbf{B}^2\rangle_\eta
&=\frac{\hbar c}{2\pi} \int_0^\infty\!\!\!\!\!d\kp\int_0^\infty\!\!\!\!\!d\kz\ \kp k  \left\{ 2+\Bigl(\frac{\kz-\kdz}{\kz+\kdz}\Bigl)^2
+\Bigl(\frac{n^2\kz-\kdz}{n^2\kz+\kdz}\Bigl)^2\right.\\
&+2 \frac{\bigl(\kp^2-\kz^2\bigr)}{k^2}\frac{\kz-\kdz}{\kz+\kdz}\cos\bigl(2\kz z\bigr)+2\frac{n^2\kz-\kdz}{n^2\kz+\kdz}\cos\bigl(2\kz z\bigr)
+\frac{\kz}{\kdz}\left[ \Bigl(\frac{2\kdz}{\kdz+\kz}\Bigr)^2
+\Bigl(\frac{2n\kdz}{\kdz+n^2\kz}\Bigr)^2\ \right]\\
&\left. +\frac{|\kz|}{\kdz} \left[\frac{4\kdz^2}{\kdz^2+|\kz|^2}
+\frac{4n^2\kdz^2}{\kdz^2+n^4|\kz|^2}\right] e^{-2|\kz|z}\right\} e^{-\eta ck}.
\end{aligned}
\end{equation}

 In the next Section we shall consider the appropriate limit of these expressions for the case of an ideal metal.

\section{\label{sec:3}The ideal conductor limit: surface divergences at the interface}

We now consider the limits $n \to 1$ and $n \to \infty$ of \eqref{EQuadroMedio}, respectively yielding the case of vacuum space and of the conductor,

\begin{equation}\label{Eqv}
\langle\mathbf{E}^2\rangle_\eta^{\mathrm{vac}}=\lim_{n\to1}\langle\mathbf{E}^2\rangle_\eta=
\frac{2\hbar c}{\pi}\int_0^\infty\!\!\!\!\! d\kp\ \int_0^\infty\!\!\!\!\! d\kz\ \kp\ \sqrt{\kp^2+\kz^2}\ e^{-c\eta\sqrt{\kp^2+\kz^2}};
\end{equation}

\begin{equation}\label{Eqc}
\langle\mathbf{E}^2\rangle_\eta^{\mathrm{con}}=\lim_{n\to\infty}\langle\mathbf{E}^2\rangle_\eta=
\frac{2\hbar c}{\pi}\int_0^\infty\!\!\!\!\! d\kp\ \int_0^\infty\!\!\!\!\! d\kz\ \frac{\kp}{\sqrt{\kp^2+\kz^2}}\ \bigl(\kp^2+2\kz^2\sin^2(\kz z)\bigr)\ e^{-c\eta\sqrt{\kp^2+\kz^2}}.
\end{equation}

Analogous expressions are obtained from \eqref{BQuadroMedio} for the magnetic fluctuations, for the vacuum space and for the conductor case,

\begin{equation}\label{Bqv}
\langle \mathbf{B}^2\rangle_\eta^{\mathrm{vac}}=\lim_{n\to1}\langle\mathbf{B}^2\rangle_\eta=
\frac{2\hbar c}{\pi}\int_0^\infty\!\!\!\!\! d\kp\ \int_0^\infty\!\!\!\!\! d\kz\ \kp\ \sqrt{\kp^2+\kz^2}\ e^{-c\eta\sqrt{\kp^2+\kz^2}};
\end{equation}

\begin{equation}\label{Bqc}
\langle\mathbf{B}^2\rangle_\eta^{\mathrm{con}}=\lim_{n\to\infty}\langle\mathbf{B}^2\rangle_\eta=
\frac{2\hbar c}{\pi}\int_0^\infty\!\!\!\!\! d\kp\ \int_0^\infty\!\!\!\!\! d\kz\  \frac{\kp}{\sqrt{\kp^2+\kz^2}}\ \bigl(\kp^2+2\kz^2\cos^2(\kz z)\bigr)\ e^{-c\eta\sqrt{\kp^2+\kz^2}}.
\end{equation}

The conductor results, both for the electric and magnetic components, can be renormalized by subtracting the spatially homogeneous vacuum contribution, that is that obtained respectively in \eqref{Eqv} and \eqref{Bqv} in the absence of the material half-space. Thus we obtain

\begin{equation}\label{Eqcr}
\langle\mathbf{E}^2\rangle_{\eta\ \mathrm{R}}^{\mathrm{con}} =
-\frac{2\hbar c}{\pi}\int_0^\infty\!\!\!\!\! d\kp\ \int_0^\infty\!\!\!\!\! d\kz\ \frac{\kp \kz^2}{\sqrt{\kp^2+\kz^2}} \cos (2\kz z)\ e^{-c\eta\sqrt{\kp^2+\kz^2}}
\end{equation}
and
\begin{equation}\label{Bqcr}
\langle\mathbf{B}^2\rangle_{\eta\ \mathrm{R}}^{\mathrm{con}} =-\langle\mathbf{E}^2\rangle_{\eta\ \mathrm{R}}^{\mathrm{con}}
\end{equation}
\end{widetext}

From the expressions above, it is evident that the time-splitting procedure introduces an exponential cut-off in the frequency integrals, suppressing contributions of modes with frequency $\omega > \wco = 1/\eta$. If $\wco$ in kept finite, the renormalized field fluctuations \eqref{Eqcr} and \eqref{Bqcr} are finite for any $z$, including at the interface $z=0$, due to the regularization introduced by the cut-off function. Also, the electric and magnetic parts are opposite each other in all points and for any value of $\eta$. If we take $\eta =0$, we get a divergence at the conductor-vacuum interface $z=0$: in this case, a cut-off function such as that given by the plasma model for a real conductor (assumed valid also at very high frequecies), that introduces a $\omega^{-2}$ factor for frequencies larger than the conductor plasma frequency, would not be enough to avoid the divergence at the interface, as showed in \cite{SF02}. There is however some controversy in the literature about which model of a real conductor (plasma, Drude or other) should be used for calculating Casimir energies and the Casimir force between real materials.

An explicit evaluation of the integrals in \eqref{Eqv}, \eqref{Eqc}, \eqref{Eqcr} yields:
\begin{equation}\label{Equadrovacuum}
\langle\mathbf{E}^2\rangle_\eta^{\mathrm{vac}}= \frac{12 \hbar }{\pi c^3 \eta ^4}
\end{equation}
for the vacuum case;
\begin{equation}
\langle\mathbf{E}^2\rangle_\eta^{\mathrm{con}}= \frac{12 \hbar }{\pi c^3 \eta ^4}+\frac{4c\hbar}{\pi} \frac{\left(12 z^2-c^2 \eta^2\right)}{\left(4 z^2+c^2 \eta ^2\right)^3}.
\end{equation}
for the conductor case, and
\begin{equation}\label{EQuadroConduttoreEta}
\Equadroeta=\frac{4c\hbar}{\pi} \frac{\left(12 z^2- c^2 \eta^2\right)}{\left(4 z^2+c^2 \eta ^2\right)^3}.
\end{equation}
for the renormalized conductor case.

As expected, $\langle\mathbf{E}^2\rangle_\eta^{\mathrm{vac}}$ and $\langle\mathbf{E}^2\rangle_\eta^{\mathrm{con}}$ diverge for $\eta \to 0$.

The limit $\eta \to 0$ of \eqref{EQuadroConduttoreEta} for $z \neq 0$ gives back the well-known result for the vacuum average value of the electric field squared in the presence of a perfectly conducting plate \cite{FS98}
\begin{equation}\label{EQuadroConduttore}
\Equadrolim=\lim_{\eta\to0}\Equadroeta=\frac{3c\hbar }{4\pi z^4} .
\end{equation}

Analogous results are obtained for the magnetic field fluctuations
\begin{equation}
\langle\mathbf{B}^2\rangle_\eta^{\mathrm{vac}}= \frac{12 \hbar }{\pi c^3 \eta ^4} ;
\end{equation}
\begin{equation}
\langle\mathbf{B}^2\rangle_\eta^{\mathrm{con}}= \frac{12 \hbar }{\pi c^3 \eta ^4}-\frac{4c\hbar}{\pi} \frac{\left(12 z^2- c^2 \eta^2\right)}{\left(4 z^2+c^2 \eta ^2\right)^3};
\end{equation}
\begin{equation}\label{BQuadroConduttoreEta}
\langle\mathbf{B}^2\rangle_{\eta\ \mathrm{R}}^{\mathrm{con}}=-\frac{4c\hbar}{\pi} \frac{\left(12 z^2- c^2 \eta^2\right)}{\left(4 z^2+c^2 \eta ^2\right)^3};
\end{equation}
\begin{equation}\label{BQuadroConduttore}
\langle\mathbf{B}^2\rangle_{\mathrm{R}}^{\mathrm{con}}=\lim_{\eta\to0}\langle\mathbf{B}^2\rangle_{\eta\ \mathrm{R.}}^{\mathrm{con}}=-\frac{3c\hbar }{4\pi z^4} .
\end{equation}

The renormalized field fluctuations $\Equadroeta$ and $\langle\mathbf{B}^2\rangle_{\eta \mathrm{R}}^{\mathrm{con}}$ given by \eqref{EQuadroConduttoreEta} and \eqref{BQuadroConduttoreEta}, which are proportional to the renormalized electric and magnetic energy densities, are finite provided $\eta$ is not vanishing.
Equations \eqref{EQuadroConduttore} and \eqref{BQuadroConduttore} show also, as it is already known, that the presence of a perfectly conducting plate increases the fluctuations of the electric field whereas reduces magnetic field fluctuations.
We can now address the main point of this paper, that is the limit to an ideal conductor, and in particular the behavior of the field fluctuations near the plate, and the presence and origin of surface divergences in our limit to the ideal conductor.
As already noticed, the use of the time-splitting procedure introduces an exponential cut-off function in the frequency integrals of Sec.\ref{sec:2} giving field fluctuations. This is equivalent to assume a cut-off function with the scale of the exponential related to a frequency such that modes of higher frequency are suppressed.
A similar cut-off function has already been used in \cite{Pfenning00} and in \cite{Milton11a} for the scalar field case.
Thus we can take expression \eqref{EQuadroConduttoreEta} and \eqref{BQuadroConduttoreEta} as the renormalized fluctuations of the electric and magnetic field for this model of a conductor with an exponential cut-off function. In this model, $1/\eta$ plays a role analogous to that of a plasma frequency (but we are not dealing with the plasma model, that yields a different form of the cut-off  function). The fact that an exponential cut-off could be not representative of a real conductor is not essential in our case, because our interest is to consider the limit $\eta \to 0$ or $\wco \to \infty$: in this limit our cut-off function is approaching one for all frequency, and we recover the physical situation of a nondispersive material. However, our procedure of considering the ideal conductor through a limiting process allows us to show the existence of surface divergences, and discuss their structure. As we shall show in the following, this does not seems possible if the field is directly quantized with the perfectly conducting plate, nor if the plasma model is used in the limit of a large plasma frequency (because, as it is shown in \cite{SF02}, the plasma model does not completely remove the divergences at the metal-vacuum interface).

We first notice that, for any $\eta >0$, the integrals of the renormalized energy densities over all vacuum space vanish for both their electric and magnetic parts, because
\begin{equation}\label{integral}
\int_0^\infty dz \frac {12 z^2-c^2\eta^2}{(4z^2+c^2\eta^2)^3}=0.
\end{equation}
Thus both electric and magnetic renormalized energy densities vanish when integrated over the $z>0$ half-space, whichever the cut-off frequency is.
This also means that, in the ideal conductor limit $\eta \to 0$, the (diverging) electric and magnetic energy for $z>0$, proportional to the spatial integrals of \eqref{EQuadroConduttore} and \eqref{BQuadroConduttore} respectively  (both behaving as $z^{-4}$ for $z \sim 0$), must be canceled by electric and magnetic energies confined at $z=0$ in the form of surface divergences at the conductor-vacuum interface. The existence of surface divergences was already guessed in \cite{Milton04} in the case of an ideal flat boundary condition for a massless scalar field; additional $\delta-$function terms were added in the Lagrangian in order to take into account the presence of the plate. A similar situation occurs for a scalar field with appropriate diverging potentials which simulate the boundary \cite{Milton11a}. Our approach of obtaining the ideal conductor through an appropriate limit process, as we will now show, actually allows us to physically understand the origin and properties of such surface divergences of the electric and magnetic energy densities, without additional hypothesis.

We now consider the behavior of the electric fluctuations and energy density for growing values of $1/\eta$, and compare it with the same quantity for an ideal metal as given by \eqref{EQuadroConduttore}.

\begin{figure}[h]\centering
\includegraphics[width=8.5cm]
{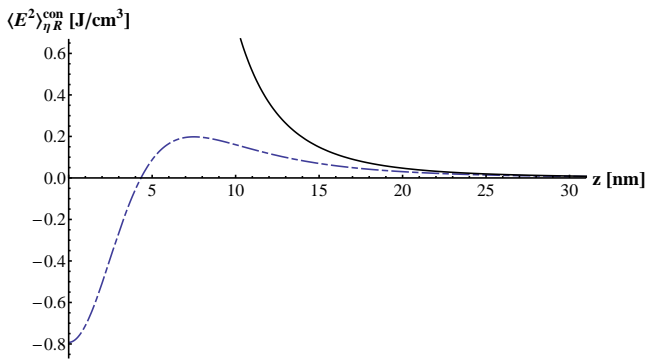}
\caption{(Color online) Comparison of $\langle\mathbf{E}^2\rangle_{\eta\ \mathrm{R}}^{\mathrm{con}}$
with $1/\eta = 2\!\times\!10^{16}$Hz
(dashed blue line) and $\Equadrolim$ (continuous black line).}\label{FigDivergences}
\end{figure}

Figure \ref{FigDivergences} shows $\langle\mathbf{E}^2\rangle_{\eta\ \mathrm{R}}^{\mathrm{con}}$ as given by \eqref{EQuadroConduttoreEta} with a value of $1/\eta$ comparable with a typical plasma frequency of a real metal, in comparison with the $z^{-4}$ behavior of $\Equadrolim$ given by \eqref{EQuadroConduttore}. Significant differences are evident in the proximity of the interface.

\begin{figure}[h]\centering
\includegraphics[width=8.7cm]
{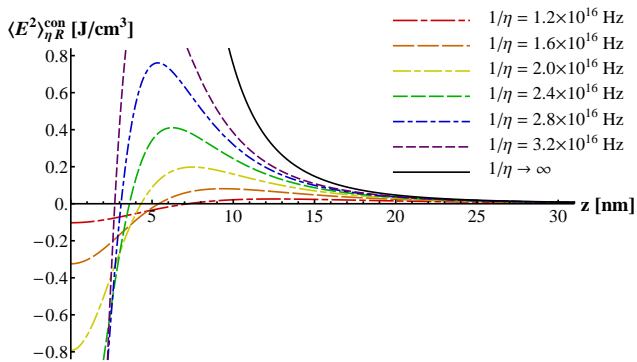}
\caption{(Color online) Comparison of $\langle\mathbf{E}^2\rangle_{\eta\ \mathrm{R}}^{\mathrm{con}}$
for different values of $\eta$. When $\eta$ decreases ($\wco$ increases), renormalized field fluctuations and energy densities for $z>0$ tend to the $z^{-4}$ law of the ideal conductor case (continuous line), whereas at $z \sim 0$ surface divergences appear with a positive and a negative peak that in the limit tend to squeeze at $z=0$.}\label{FigComDivergences}
\end{figure}

Figure \ref{FigComDivergences} shows the behavior of $\langle\mathbf{E}^2\rangle_{\eta\ \mathrm{R}}^{\mathrm{con}}$ for different values of $\eta$, and compared with $\Equadrolim$ (continuous line). For any nonvanishing value of $\eta$, i.e. for any finite value of the cut-ff frequency $\wco$, renormalized fluctuations and energy densities are finite in all points of the vacuum region and there are not divergences at the surface; $\langle\mathbf{E}^2\rangle_{\eta\ \mathrm{R.}}^{\mathrm{con}}$ has a maximum at $z_\eta^{\mathrm{max}}=\frac{\eta c}{2}$ with the positive value $\frac{c\hbar}{\pi}\frac{1}{\eta^4c^4}$, and a minimum at $z_\eta^{\mathrm{min}}=0$ with the negative value  $-\frac{c\hbar}{\pi}\frac{4}{\eta^4c^4}$. The width of the curve, that can be estimated as the distance between the two inflection points around the maximum, is $\Delta_\eta \simeq\!0.5\eta c$.
For an increasing cut-off frequency $1/\eta$, the curves in figure approach the ideal conductor limit for large distances from the interface, but significant differences still remain close to the surface. In fact, while the ideal conductor limit $\Equadrolim$ diverges with positive values at the surface, $\langle\mathbf{E}^2\rangle_{\eta\ \mathrm{R}}^{\mathrm{con}}$ assumes more and more negative values as $1/\eta \to \infty$, and the width of the curve reduces to zero. Maximum and minimum values of $\langle\mathbf{E}^2\rangle_{\eta\ \mathrm{R}}^{\mathrm{con}}$ tend to collapse each other at the surface in the ideal conductor limit, yielding a surface divergence containing a nonvanishing electric and magnetic energy
\begin{equation}
 z_\eta^{\mathrm{max}}\xrightarrow[\eta\to0]{} z_\eta^{\mathrm{min}}=0.
\end{equation}
Thus, by analyzing the limiting case $\wco \to \infty$ ($\eta \to 0$) it becomes evident that the well-known diverging behavior as $z^{-4}$ near the surface for the ideal conductor, indeed originates from the maximum of $\langle\mathbf{E}^2\rangle_{\eta\ \mathrm{R}}^{\mathrm{con}}$. This is clearly shown in Figure \ref{FigComDivergences}. The negative divergence at $z=0$ when $\eta \to 0$ is completely lost if the calculation is directly performed for an ideal conductor. Our approach thus makes clear the origin of the surface divergences of the renormalized squares of electric and magnetic field; these divergences, and their detailed structure, naturally appear in the limit process from the dielectric to the ideal conductor that we have used.

The limit $\wco \to \infty$ is well defined in our model for any distance $z$ except $z=0$. Also, due to \eqref{integral}, both integrals over all vacuum space of the (renormalized) electric and magnetic energy densities vanish, as expected from the evaluation of global field energies. This feature does not seem to occur if the calculation is performed directly for an ideal boundary or the plasma model is used even in the limit of an infinite plasma frequency (see \cite{SF02}), because in these cases the surface divergences are not fully included in the expressions of the field energy densities.
By taking into account the surface divergences with our limit procedure, we thus obtain consistency between global field energies and (integrated) local energy densities.

Field fluctuations and energy densities near the surface can be investigated through the retarded Casimir-Polder interaction energy with an appropriate polarizable body with static polarizability $\alpha$. When the distance $d$ between the surface and this body is larger that the wavelength associated to its main transition, the so-called far zone approximation holds; in this case the Casimir-Polder energy is given by $\Delta E_E = -\alpha \langle\mathbf{E}^2\rangle_{\eta\ \mathrm{R}}^{\mathrm{con}} (d)/2$ (see for example \cite{PPT98}). A similar expression holds for the magnetic fluctuations.

\section{\label{sec:4}Conclusions}

In this paper we have considered zero-point electric and magnetic field fluctuations (or equivalently field energy densities) in the vacuum space, when half-space is filled with a homogeneous non-dissipative dielectric. The material is characterized by a constant real refractive index $n$. We have then taken the limit $n \to \infty$ and, by introducing a time-splitting procedure with a parameter $\eta$, we have mathematically included also a high-frequency exponential cut-off function characterized by a cut-off frequency $\wco = 1/\eta$. We have considered in detail two successive limits: $n \to \infty$ with $\wco$ finite, and then $\wco \to \infty$ (ideal conductor limit). We have found that no divergences in the renormalized field fluctuations (and in the renormalized field energy densities) are present if $\wco$ is kept finite. In this case, field fluctuations at small distances from the interface significantly differ compared to known results for a perfectly conducting plate, as shown in Figures \ref{FigDivergences} and \ref{FigComDivergences}. In the limit $\wco \to \infty$ (ideal conductor limit) surface divergences of the field fluctuations (or energy densities) at the interface $z=0$ are found, while for $z > 0$ the fluctuations approach the well-known $z^{-4}$ behavior of the perfect conductor case. The structure of the surface divergences has been discussed in detail. These surface divergences of field fluctuations are not obtained if the field is directly quantized in the presence of an ideal metallic surface; also the plasma model does not seem to allow to obtain their structure. Thus our approach of obtaining the ideal metal through an appropriate limit procedure starting from a dielectric has allowed us to obtain in a natural way the explicit structure of the surface divergence at the interface between vacuum and a conducting material. Having taken into account the field energy in the surface divergences has also allowed us to show consistence between global and local field energies. Finally, we have stressed that field fluctuations and energy densities near the interface can in principle be investigated through the Casimir-Polder interaction energy with an appropriate electrically or magnetically polarizable body placed near the interface.

\begin{acknowledgments}
The authors acknowledge support from the ESF Research Networking Program CASIMIR. Financial support by Ministero dell'Istruzione, dell'Universit\`{a} e della Ricerca and by Comitato Regionale di Ricerche Nucleari e di Struttura della Materia is also acknowledged.
\end{acknowledgments}

\end{document}